\documentclass[a4paper,11pt]{article}
\usepackage{pos}
\usepackage{tabu}

\newcolumntype{M}[1]{>{\centering\arraybackslash}m{#1}}
\newcommand{\mycite}[1]{\,\cite{#1}}

\newcommand{\hu}{\,\mathrm{km\,s^{-1}Mpc^{-1}}}

\newcommand{\geff}{G_{\rm eff}}

\newcommand{\lgeff}{{\rm log}_{10}(G_{\rm eff}/{\rm MeV^{-2}})}

\newcommand{\lcdm}{$\Lambda$CDM}
\newcommand{\sinu}{SINU}
\newcommand{\three}{$\mathbf {3c+0f}$}
\newcommand{\two}{$\mathbf {2c+1f}$}
\newcommand{\one}{$\mathbf {1c+2f}$}

\newcommand{\cb}[1]{\textcolor{black}{#1}}

\title{Self-interacting neutrinos as a solution to the Hubble tension?}

\author*[a]{Anirban Das}
\author[b]{Subhajit Ghosh}

\affiliation[a]{SLAC National Accelerator Laboratory,\\
  2575 Sand Hill Road, Menlo Park, California 94025, USA}

\affiliation[b]{Department of Physics, University of Notre Dame,\\
	 South Bend, IN 46556, USA}

\emailAdd{anirband@slac.stanford.edu}
\emailAdd{sghosh5@nd.edu}

\abstract{Self-interaction among the neutrinos in the early Universe has been proposed as a solution to the Hubble tension, a discrepancy between the measured values of the Hubble constant from CMB and low-redshift data. However, flavor-universal neutrino self-interaction is highly constrained by Big Bang Nucleosynthesis and other laboratory experiments such as, tau and K-meson decay, double-neutrino beta decay etc. We study the cosmology when only one or two neutrino states are self-interacting. Such flavor-specific interactions are less constrained by the laboratory experiments. Lastly, we address the feasibility of resolving the Hubble tension within the framework of such flavor-specific neutrino self-interaction.}

\FullConference{%
  *** The European Physical Society Conference on High Energy Physics (EPS-HEP2021), ***\\
  *** 26-30 July 2021 ***\\
  *** Online conference, jointly organized by Universität Hamburg and the research center DESY ***
}

\begin{document}
\maketitle

\section{Introduction}
Since the early days of the Planck experiment, a \mbox{discrepancy} in the value of the Hubble parameter from CMB observations and local (low redshift) measurements have become a topic of active research\mycite{Ade:2013zuv}. This discrepancy has been heightened further with the publication of the latest data from the Planck experiment and the local $H_0$-measurement using cepheids and supernovae by the SH0ES collaboration\mycite{Aghanim:2018eyx,Riess:2019cxk,2018ApJ...855..136R}. Planck measured value a Hubble constant using the CMB data $H_0=67.36\pm 0.54\hu$, and the SH0ES collaboration measured $H_0=74.03\pm 1.42\hu$. There is now a $4.4\sigma$ tension between these two values. An independent local measurement of $H_0$ performed in Ref.\cite{freedman2019} found $H_0=69.8\pm 1.9\hu$. Other than some unknown experimental systematics, this discrepancy could imply a new physics beyond \lcdm. 
Several such scenarios have been proposed to address it. See Ref.\,\mycite{DiValentino:2021izs} for a review. 

The authors in Ref.\mycite{Cyr-Racine:2013jua} introduced a model of self-interacting neutrinos (\sinu) where the interaction is mediated by a heavy scalar $\phi$.  At low temperature, it leads to a flavor-universal four-Fermi interaction with a coupling strength $\geff$.
A Bayesian analysis of this model with the CMB data yields two modes in the posterior distribution of $\geff$: a strongly interacting (SI) mode with a large value $\lgeff \simeq -1.711^{+0.099}_{-0.11}$ ($68\%$ confidence limit), and a moderately interacting (MI) mode with an upper bound $\lgeff < -3.57$ at $95\%$ confidence level when the Planck 2015 and BAO data are used. 
However, \sinu{} scenario faces strong constraints from the laboratory experiments \mycite{Blinov:2019gcj,Lyu:2020lps}. They showed that a simple model of flavor-universal \sinu{} scenario is ruled out by laboratory constraints from meson decay, $ \tau $ decay, and double beta decay.

In Ref.\,\cite{Das:2020xke}, we performed a Bayesian analysis of flavor-specific neutrino self-interaction scenario using latest cosmological data. Our goal was to complement the flavor-specific \sinu{} studies from laboratory experiments with the cosmological data. 
We considered three massless neutrinos and fixed $ N_{\rm eff}  = 3.046$. The effect of neutrino mass can be speculated from our results. We consider three scenarios depending on the number of interacting species -- 3-coupled (\three), 2-coupled + 1-free-streaming (\two), and 1-coupled + 2-free-streaming (\one) respectively. We assumed same coupling strengths for the coupled species for the first two cases. 

\section{The model}\label{sec:model}
We consider scalar interactions between massless $\nu$ and $\phi$ as
\begin{equation}
	\mathcal{L} \supset g_{ij}\phi\bar{\nu}_i\nu_j\,,
\end{equation}
where $g_{ij}$ is the coupling between $\phi$ and the neutrino flavors $i$ and $j$. At low temperature, a four-Fermi interaction among the neutrinos is generated as follows,
\begin{equation}\label{eq:geff2}
	\mathcal{L} \supset \geff^{(ijkl)}\bar{\nu}_i\nu_j\bar{\nu}_k\nu_l, \qquad \geff^{(ijkl)} \equiv \frac{g_{ij}g_{kl}}{M_\phi^2}\,.
\end{equation}
We consider only \emph{diagonal} interactions in the flavor space.
The number of neutrino flavors will be fixed to three with all of them having a same temperature $T_\nu$. However, because of the complete equivalence of the interacting states in the context of CMB, we need to consider only one common coupling parameter $\geff$ for all the interacting states in a given scenario. The \cb{comoving} neutrino self-interaction opacity $\dot{\tau}_\nu$ is defined as
\begin{equation}\label{eq:opacity}
	\dot{\tau}_\nu = -a(\geff)^2T_\nu^5\,,
\end{equation}
where $a$ is the scale factor of the Universe. 
\begin{figure}[t]
	\centering
	\includegraphics[width=0.87\linewidth]{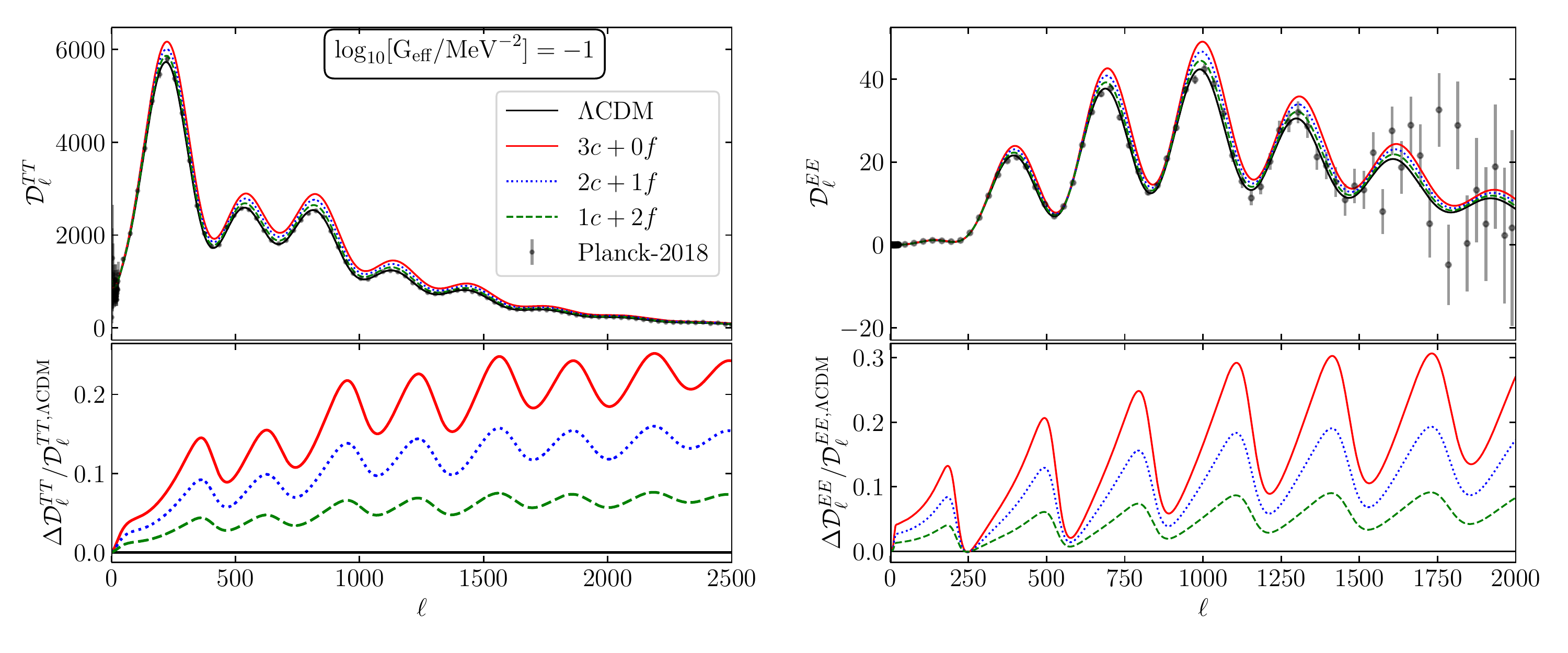}
	\caption{CMB $ TT $ and $ EE $ angular power spectra for \three{} (red, solid), \two{} (blue, dotted), and \one{} (green, dashed) scenarios  are shown in the top panels. The \lcdm{} power spectra are also shown for comparison in solid black. The parameters for the \lcdm{} spectra correspond to the best-fit points for the TT,TE,EE+lowE dataset. The bottom panels show the relative changes from the \lcdm{} spectra. For \sinu{} plots, we have set $ \lgeff =-1 $ and the rest of the parameters are fixed to their \lcdm{} best-fit values. We also show the binned Planck 2018 data in both the plots as black circles with errorbar.}\label{fig:cmb_spectra}
\end{figure}

\section{Changes in the CMB power spectra}\label{sec:changes_in_CMB}
The self-interaction stops the neutrinos from free-streaming before decoupling. The new interaction damps the perturbations for $\ell \ge 2$ and impedes the growth of the anisotropic stress $\sigma$. The damping is maximum for \three{} where all three neutrinos are interacting, and gradually decreases for \two{} and \one{} where the number of interacting neutrino flavor is two and one respectively. The anisotropic stress is related to the gravitational potentials $\phi$ and $\psi$ via
\begin{equation}
	k^2(\phi-\psi) = 12\pi Ga^2\sum_{i=\gamma,\nu}(\rho_i+P_i)\sigma_i\, \simeq 16\pi Ga^2\rho_{\rm tot}R_\nu\sigma_\nu\,,\quad R_\nu = \dfrac{\rho_\nu}{\rho_\nu + \rho_\gamma}\;.
\end{equation}
In the last step in the equation, we ignored the small anisotropic stress of photon $\sigma_\gamma$ before recombination. And, $R_\nu\simeq 0.41$ is the fractional energy density of \emph{free streaming} neutrinos which, in radiation domination in \lcdm. The suppression of neutrino anisotropic stress helps enhance the gravitational potentials $\phi$ and $\psi$. The gravitational potentials in turn affect the evolution of the photon perturbations.

In \lcdm{} cosmology, the supersonic propagation of neutrino perturbations creates a phase shift $\phi_\nu$ and an amplitude modification $\Delta_\nu$ in the acoustic oscillations of the photon\,\cite{Bashinsky:2003tk}\,, which in the radiation domination, are given by
\begin{equation}\label{eq:phase_shift}
	\phi_\nu \simeq 0.19\pi R_\nu\,,\quad \Delta_\nu \simeq -0.27 R_\nu\,.
\end{equation}
Together, they move the peaks in the TT and EE power spectra towards smaller $\ell$, and suppress the amplitude of the power spectra. The self-interaction stops the neutrinos from free-streaming and delays the neutrino decoupling. As a result, the free-streaming neutrino fraction $R_\nu$ is decreased relative to its \lcdm{} value depending on the number of neutrino species which are coupled.

\section{Results}\label{sec:result}
The posterior of $\geff$ shows a bimodal feature as expected. The MI mode corresponds to smaller values of $\geff$ where the CMB spectra are compatible with \lcdm. The SI mode is characterized by a relatively large value of $\geff$, and arises due to degeneracy with other parameters, including $H_0$. The significance of the SI mode increases as the number of interacting neutrinos goes down. This follows from the fact that the changes in the CMB spectra in \sinu{} are proportional to the number of interacting neutrinos as can be seen from Eq.(\ref{eq:phase_shift}). This makes the SI mode parameter values more compatible with data, and boosts the significance of the mode.
\begin{figure}[t]
	\centering
	\includegraphics[width=0.42\linewidth]{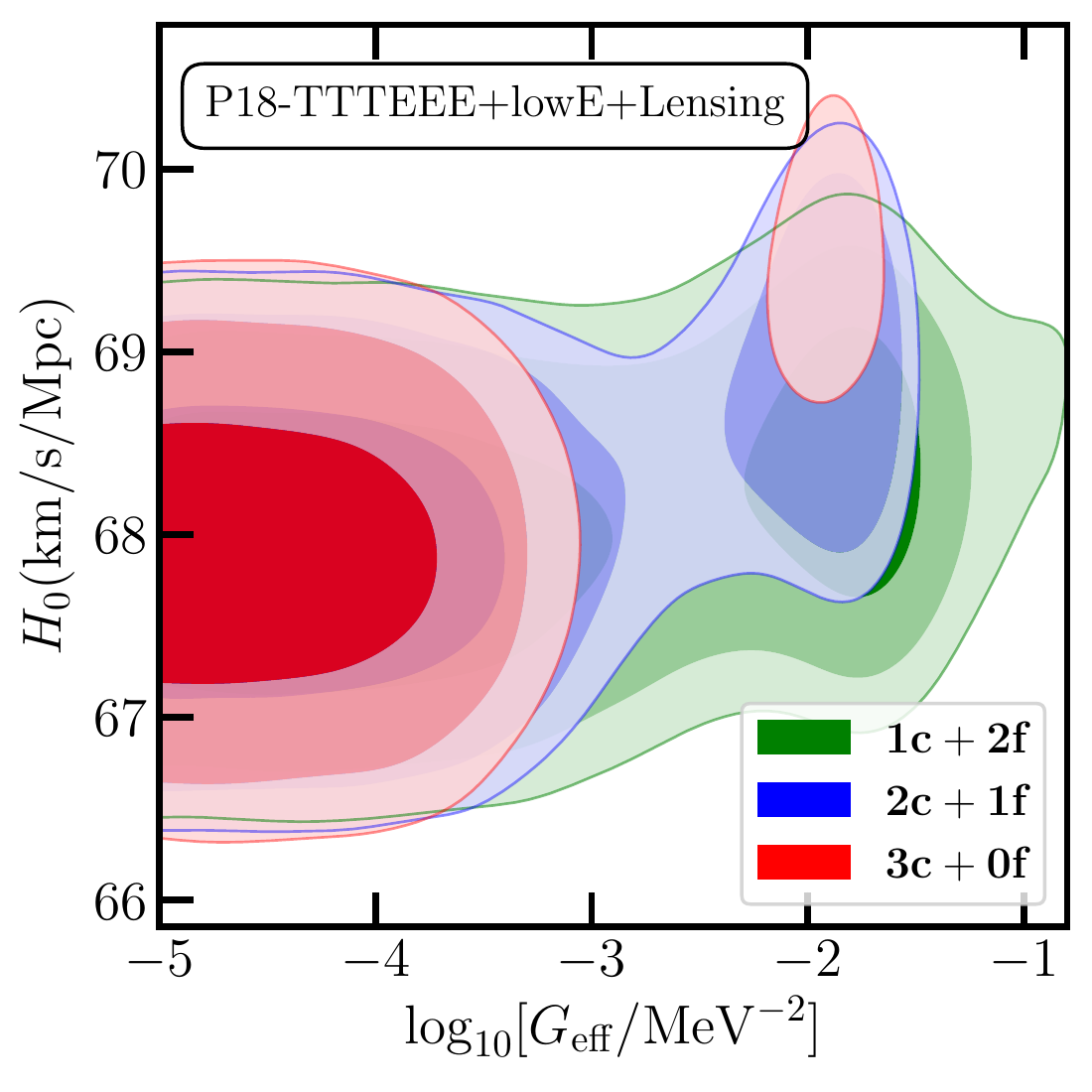}
	\caption{The contours of 68\%, 95\%, and 99\% confidence levels for $\lgeff$ and $H_0$ in \one{} (green), \two{} (blue), and \three{} (red) for the Planck TTTEEE+lowE+lensing dataset. The SI mode contour for \three{} yields the largest value of $H_0$ because of bigger phase shift. However, the significance of the mode is much less compared to \two{} and \one{}.}
	\label{fig:2d-h0}
\end{figure}

The neutrino self-coupling strength $\geff$ has a positive correlation with $H_0$ through the phase-shift. The mean value of $H_0$ shifts to $H_0=69.46\pm0.52\hu$ for the TTTEEE+lowE dataset for the SI mode, reducing the tension with the SH0ES measurement to $\sim3\sigma$. On the other hand, this value of $H_0$ is fully consistent with the CCHP measurement $H_0=69.8\pm 1.9\hu$\,\cite{freedman2019}. However, the SI mode value of $H_0$ slightly decreases in the flavor-specific \two{} and \one{} scenarios because of smaller phase shift, but the significance of the mode is increased. This is evident from figure~\ref{fig:2d-h0}. In table~\ref{tab:h0}, we show $H_0$, $\Omega_\Lambda$, $100\theta_s$, $r_s^\ast$, and $D_A^\ast$ for \three{}, \two{} and \lcdm. The shift in $\theta_s$ due to the phase shift in \sinu{} decreases $D_A^\ast$ that helps increase the value of $H_0$ and $\Omega_\Lambda$. Note that, the value of $r_s^\ast$ changes by  only about $1\sigma$ from \lcdm, and plays a sub-dominant role in changing the value of $H_0$.

\section{Discussion}
Neutrino self-interaction introduces a \emph{phase shift} and \emph{enhance} the CMB anisotropy power spectra relative to \lcdm. This scenario has been proposed to ease the Hubble tension. However, the required strength of the self-interaction to resolve the tension is strongly constrained in flavor-universal scenario by other laboratory experiments. We studied the cases where only two or one of the three neutrino species are interacting using the latest CMB data from 2018 Planck experiment. We found a bimodal posterior as in the previous studies with flavor-universal case. Moreover, we found that the strongly interacting (SI) mode is boosted in the flavor-specific scenario. Less number of interacting neutrino means less amount of changes in the CMB power spectra which makes larger coupling strength more compatible with the data boosting the SI mode.
\begin{table}[t]
	\centering
	\begin{tabular}{c|M{3cm}|M{3cm}|M{3cm}}
		\hline
		\hline
		& SI: \three& SI: \two & \lcdm{}\\
		\hline
		$ H_0 (\hu)$ &$ 69.47\pm 0.59 $ &$ 68.87\pm0.58 $ &$67.90\pm0.54$\\[1ex]
		$\Omega_\Lambda$&$ 0.7035\pm0.0071 $&$ 0.6989\pm0.0072 $&$ 0.6912 \pm 0.0073 $\\[1ex]
		$ 100\theta_s $&$ 1.0463\pm 0.00094$&$ 1.0447 \pm 0.00079 $&$1.04186\pm 0.00029 $\\[1ex]
		$r_s^\ast({\rm Mpc})$&$ 144.58\pm0.32 $&$ 144.69 \pm 0.31 $&$ 144.87\pm0.29 $\\[1ex]
		$D_A^\ast({\rm Mpc})$&$ 12.69\pm 0.036 $&$ 12.72 \pm 0.034$&$ 12.773\pm 0.028$\\[1ex]
		\hline
	\end{tabular}
\caption{Parameter values and 68\% confidence limits for SI mode in \three{} and \two, and \lcdm{} in TTTEEE+lowE+lensing data.}\label{tab:h0}
\end{table}

The neutrino coupling strength $\geff$ is degenerate with the Hubble constant $H_0$, and the SI mode favors a slightly larger value of $H_0$ compared to \lcdm, easing the Hubble tension. We, however, found that the value of $H_0$ is decreased when only two or one neutrinos are interacting. As a result, the \sinu{} scenario cannot resolve the Hubble tension when all laboratory and cosmological constraints are taken into consideration. In our work, we did not include off-diagonal interactions as it requires one to incorporate neutrino mass as well. Hence, it remains to be seen how the addition of off-diagonal interaction affects the Hubble parameter.

\acknowledgments
AD was supported by the U.S. Department of Energy under contract number DE-AC02-76SF00515. SG was supported in part by the National Science Foundation under Grant Number PHY-2014165 and PHY-1820860. SG also acknowledges the support from Department of Atomic Energy, Government of India during the initial stage of this project. This work used the computational facility of Department of Theoretical Physics, Tata Institute of Fundamental Research.

\bibliographystyle{jhep}
\bibliography{sinu}



\end{document}